\documentclass[fleqn,usenatbib]{mnras}

\usepackage{newtxtext,newtxmath}

\usepackage[T1]{fontenc}
\usepackage{ae,aecompl}
\usepackage[normalem]{ulem}

\usepackage{graphicx}	
\usepackage{amsmath}	
\usepackage{amssymb}	

\title[Photopolarimetry of AR Sco]{Time series photopolarimetry and modelling of the white dwarf pulsar in AR Scorpii}

\author[S. B. Potter et al.]{
Stephen. B. Potter,$^{1}$\thanks{E-mail: sbp@saao.ac.za}
and  David A. H. Buckley$^{1}$
\\
$^{1}$South African Astronomical Observatory, PO Box 9, Observatory, 7935, Cape Town, South Africa\\
}

\date{Accepted XXX. Received YYY; in original form ZZZ}

\pubyear{2017}

\begin{document}
\label{firstpage}
\pagerange{\pageref{firstpage}--\pageref{lastpage}}
\maketitle

\begin{abstract}
We present detailed optical photo-polarimetric observations of the recently-discovered white dwarf pulsar AR Scorpii. Our extensive dataset reveals that the polarized emission is remarkably stable and repeatable with spin, beat and orbital modulations. This has enabled us to construct a simple geometrical model which assumes that all of the optically polarized emission emanates from two diametrically opposed synchrotron emission regions on the white dwarf magnetosphere. We suggest that the observed polarimetric modulations occur as a result of an enhanced injection of relativistic electrons into the magnetosphere of the white dwarf as it sweeps past the M-dwarf. This leads to an increase in synchrotron emission as the injected electrons accelerate towards each magnetic mirror point close to the magnetic poles of the white dwarf. Whilst this scenario reproduces the detailed polarimetric modulations, other suggested scenarios involving emission sites locked in the white dwarf rotating frame are not ruled out. For example, pulsar-like particle acceleration as a result of either electric potentials between the white dwarf and the light cylinder or a striped relativistic magnetohydrodynamic wind, outside the light cylinder. Previous conclusions that argued that the observed strong optical beat modulations require that the optical polarization arises predominantly from or near the M-dwarf are inconsistent with our observations.

\end{abstract}




\section{Introduction}

The discovery of pulsar-like behaviour in the close binary system AR Scorpii (hereafter AR Sco) has firmly demonstrated that white dwarfs can exhibit many of the same characteristics as neutron star pulsars. \cite{Marsh2016} were the first to discover highly pulsed (up to 90\% pulse fraction) non-thermal emission from the spinning white dwarf in AR Sco, over a wavelength range extending from radio to the ultraviolet. These pulsations are seen predominantly at the 118 s beat period between the 117 s spin period and the 3.6 h orbital period of the binary. Perhaps the most intriguing aspect was that \cite{Marsh2016} concluded that the bulk of the luminosity of the system was a result of spin-down power of the white dwarf, for which they measured a spin-down of $\dot{P} = 3.92 \times 10^{-13}  \, \mbox{s s}^{-1}$. However a recent reevaluation \citep{Potter2018} of the spin ephemeris shows that the proposed spin-down ephemeris derived by \cite{Marsh2016} is inconsistent with newer and higher cadence observations over a 2 year timebase, which formally requires no $\dot P$ term. This implies that the spin-evolution of the white dwarf is still an open question, in particular what the actual value of spin-down power is.

The overall spectral energy distribution of AR Sco was shown by \cite{Marsh2016} to be dominated by two non-thermal power law ($ S_{\nu} \propto \nu^{\alpha}$) components, although the thermal contribution from the M5 companion is also evident. From radio to infrared frequencies ($\nu \leq 10^{12} - 10^{13} \, \mbox{Hz}$) the power law slope is $\alpha \sim 1.3$, typical of self absorbed synchrotron emission, while for higher optical to X-ray frequencies ($\nu \geq \mbox{few} \times 10^{14} \, \mbox{Hz}$),  $\alpha \sim -0.2$ \citep{Marsh2016, Geng2016}. However, the highest frequency (X-ray) constraint was quite loose, based only a short {\it Swift} ToO observation. More recent observations of AR Sco have extended the wavelength range and/or time coverage and sensitivity. X-ray observations with {\it XMM-Newton}, reported by \cite{takata2017}, shows both orbital and beat variations and an X-ray spectrum characterized by a hot multi-temperature thermal plasma ($kT \sim 1-8$ keV, \cite{takata2017}). \cite{Marcote2017} and \cite{Stanway2018} have presented results of radio observations of AR Sco, namely high resolution interferometry and time resolved eVLA observations, respectively. The source was unresolved in the former observation, while strong modulations of the radio flux on the orbital and beat periods were reported for the latter.  Long cadence optical photometry, utilizing {\it Kepler} data, has been reported in \cite{Littlefield2017}.

The strong spin modulated emission due to the white dwarf plus the dominant non-thermal nature of the SED led to the initial conclusions that magnetic interactions were powering the emission in AR Sco \citep{Marsh2016, Geng2016, Katz2017}, in many respects resembling the emission from pulsars. 
Further evidence of pulsar behaviour followed from photopolarimetric observations by \citet{Buckley2017}, who detected spin and beat modulated linear polarization as high as $40\%$. They interpreted the characteristics of the polarized and non-polarized emission in terms of synchrotron emission from two different regions, one associated with the rotating white dwarf magnetic field and the other as a result of MHD interactions with the M-dwarf companion. A number of models have now been suggested to explain the observed behaviour of AR Sco \citep{Marsh2016, Geng2016, Katz2017, Buckley2017, takata2017, takata2018}, based on some kind of white dwarf$-$red dwarf interaction, resulting in emission from the surface or coronal loops of the companion star, its magnetosphere or possibly through an associated bow shock \citep{Geng2016, Katz2017}.

The eVLA results reported in \cite{Stanway2018} showed strong beat-phase pulsations at 9 GHz, decreasing in strength with decreasing frequency. Unlike the optical polarized emission \citep{Buckley2017}, the radio emission shows only weak linear polarization but very strong circular polarization, reaching $\sim$30$\%$. \cite{Stanway2018} postulate the existence of a non-relativistic cyclotron emission component, dominating at radio frequencies, which given the likely magnetic field strength of the emission region, arises in the vicinity of the M-dwarf.

In this paper we present new extensive optical polarimetric results from $\sim$65 h of observations over a period of two consecutive years. We derive a high quality high resolution periodogram for the polarimetric variations which, although appearing to be quite complex, is entirely explained by the presence of just two basic periodicities, namely the spin frequency ($\omega$) and the orbital frequency ($\Omega$), together with the sideband frequencies ($\nu = m.\omega \pm n.\Omega$, where {\it m, n} are integers), their aliases and harmonics. We present a simple geometric model based on synchrotron emission in the magnetosphere of the white dwarf which can qualitatively explain the polarized emission as a function of orbital, spin and beat phase.

\section{Observations}

In Table 1 we show a log of all the observations of AR Sco which were made with the \textit{HI-speed Photo-POlarimeter} (HIPPO; \cite{potter2010}) on the 1.9-m telescope of the South African Astronomical Observatory. We have already reported on the high speed photometry results obtained during this observing campaign \citep{Potter2018}. This paper can be consulted for details of the instrument and observational setup. In summary, HIPPO was operated in ``all-stokes" mode,  simultaneously measuring all four Stokes parameters ({\it Q, U, V, I}) at an intrinsic time resolution of 0.1 s. These data were then binned to a time resolution of 10 s before undertaking further analysis.

Several polarized and non-polarized standard stars \citep{Hsu1982, bastien1988} were observed in order to calculate the position angle offsets, instrumental polarization and efficiency factors. Photometric calibrations were not carried out; photometry is given as total counts. Background sky polarization measurements were taken at frequent intervals during the observations.

\begin{table}
	\centering
	\caption{Table of observations. All observations were made with the HIgh-speed-Photo-Polarimeter \citep{potter2010} on the SAAO 1.9m telescope. The March 2016 observations are from \citep{Buckley2017}}.
	\label{tab:example_table}
	\begin{tabular}{lccr} 
		\hline
		Date & No.Hours & Filter(s) \\
		\hline
		14 Mar 2016 & 0.57 & OG570, clear \\
		15 Mar 2016 & 1.68 & OG570, clear \\
        14 May 2016 & 5.6 & clear \\
		15 May 2016 & 7.8 & OG570 \\
		16 May 2016 & 6.28 & I \\
		25 May 2016 & 4.7 & OG570, clear \\
		26 May 2016 & 7.37 & OG570, clear \\
		27 May 2016 & 6.82 & OG570, clear \\
		28 May 2016 & 5.98 & OG570, clear \\
		22 Mar 2017 & 1.68 & OG570 \\
		27 Mar 2017 & 4.03 & clear \\
		28 Mar 2017 & 4.07 & clear \\
        23 Jun 2017 & 8 & clear \\
		\hline	
	\end{tabular}
\end{table}

\begin{figure*}
    \centering
	\includegraphics[width=\textwidth]{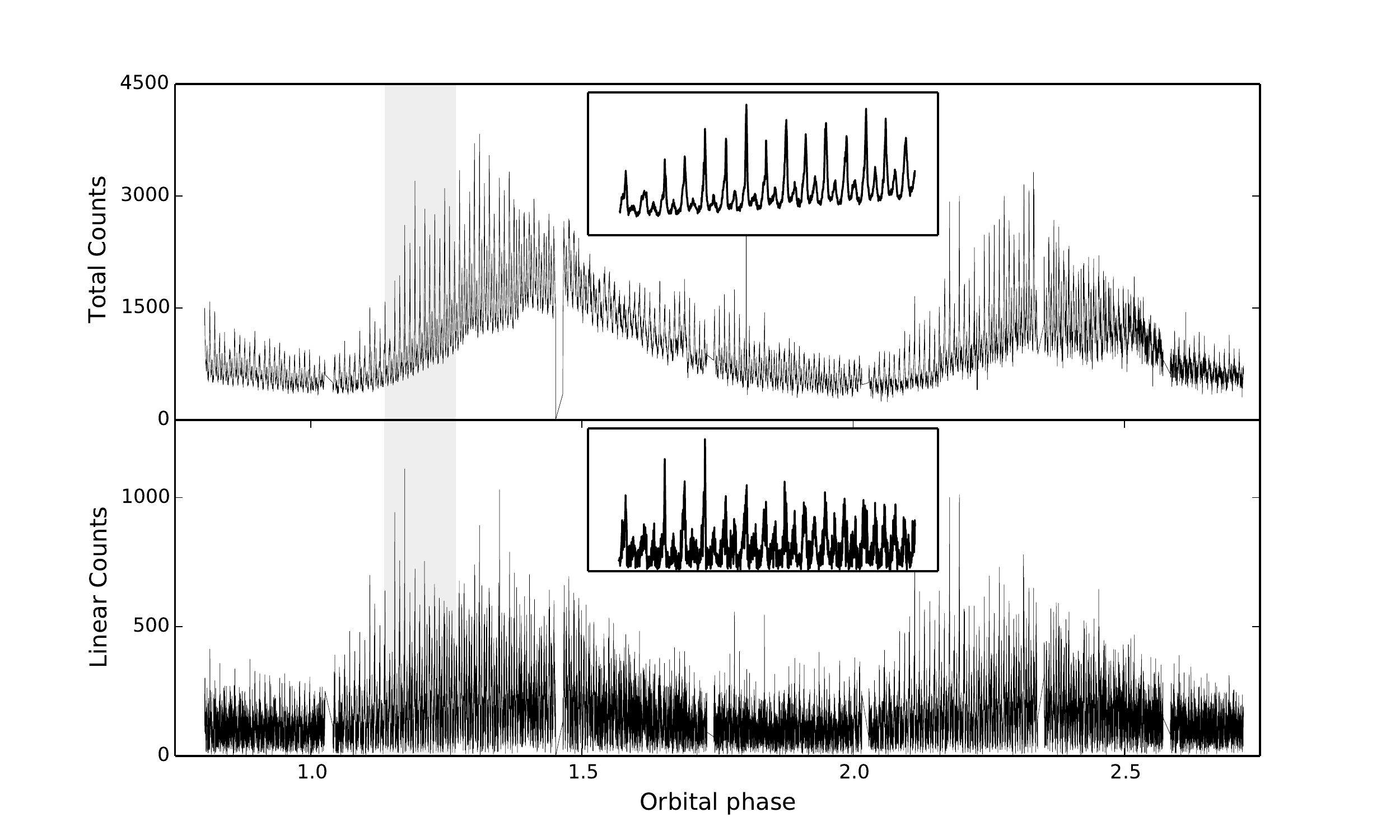}
    \caption{An example dataset, specifically the $\sim$6.8 hours, clear filtered photo-polarimetric observations of 27 May 2016. Top and bottom panels are the photometry and total linear counts respectively. The spin and beat pulses can clearly be seen in the insets which correspond to $\sim$30 minute expanded views, indicated by the grey regions.}
        \label{fig:example_figure}
\end{figure*}

\section{The photo-polarimetry}
In Fig. 1 we present an example of one of our observations, namely on the night of 2016 May 27 taken with the clear filter. The top and bottom panels show, respectively,  the simultaneous photometry (presented in \cite{Potter2018}) and linearly polarized counts. In both plots of Fig. 1 we see both the low frequency high amplitude orbital modulation, covering $\sim$2 cycles, plus the strong beat and spin pulses at a period of $\sim$2 minutes and the first harmonic at half this period. We also show in the figure insets expanded sections of the observations, which present spin/beat modulations in more detail. Both the photometry and polarimetry are consistent with the results reported by \cite{Marsh2016} and \cite{Buckley2017}, respectively, although those data do not cover entire orbital cycles.

In \cite{Potter2018} we presented a method to graphically demonstrate how the beat and spin variations are modulated on the orbital period. In Fig.2 we show a very similar plot to one presented in that paper, but extended to show the details of the spin and beat modulated polarization behaviour.  In summary, these are in the form of a 2D colour coded image of the pulse profiles, where the details of how they were derived are presented in \cite{Potter2018}. We refer to these henceforth as ``dynamic pulse profiles''. The top two panels in Fig. 2, which show the photometric observations, are a reproduction of Figure 2 of \cite{Potter2018} and are included here to allow easy comparison between the total flux and the polarized flux behaviour. These dynamic pulse profile images are extremely stable over time and show exactly the same morphology from one orbital cycle to another, even if separated by over 1 year. Therefore in order to improve the signal to noise of these images, we have combined all of our white-light (``clear'') observations, which is by far the largest dataset, to produce the final version of Fig. 2.

The dynamic pulse profiles show that the double-peaked spin and beat pulses evolve in amplitude over the orbital cycle peaking at $\sim$0.4$-$0.5 in orbital phase and significantly reduced at orbital phase $\sim$0. There also appears to be a second set of double pulses (spin and beat) between orbital phases $\sim$0.6$-$1.0. The spin and beat pulses also appear not to be stable in phase, i.e. the spin and beat pulses appear to drift later and earlier respectively as a function of orbital phase, giving the diagonal appearance. The "slopes" of the diagonal pulses in the spin/beat-orbit phase-space is consistent with cross "contamination" between the spin and beat frequencies.

The second two panels of Fig. 2 show the simultaneous linearly polarized flux. These were constructed by phase-fold binning in the same manner as the photometric flux. The linear polarization displays similar double peaked-spin and beat amplitude variations to the photometry. In particular the amplitude of the peaks are strongest around orbital phases $\sim$0.2$-$0.5 and the spin and beat pulses appear to drift later and earlier respectively as a function of orbital phase, giving the diagonal appearance. However, the linear pulses also appear to be doubled compared to the photometric pulses. e.g. the main photometric spin pulse centered on orbital/spin phase $\sim$0.35/0.15 appears diagonally split into two at the corresponding location in the linear polarization image. The second smaller photometric peak centered on orbital/spin phase $\sim$0.4/0.65 is similarly diagonally split into two linear polarization peaks. 

The beat pulses are likewise split, but more in the vertical (orbital) direction. A close inspection of the orbit-beat phase linear polarization image one can see a ``darker'' diagonal band running from the top left to the bottom right and passing through the center of the fainter of the ``split'' peaks. A less obvious parallel band dark band can be seen passing through the center of the brighter split peak. These darker bands are indicative of a minimum non-zero amount of polarized flux at these phases.

The third row of panels in Fig. 2 show the simultaneous circularly polarized flux also phase-folded and binned in the same manner as the photometry and linear polarization. The amount of circular polarization is at a much lower level than the linear polarization. Individual light curves (not shown) show only occasional seemingly random excursions of circular polarization, peaking at values of $\sim$3 percent. The co-adding of the multiple data sets reveals that there are regions in the orbit, spin/beat phase space where circular polarization is preferentially seen. Specifically, positive and negative circular polarization (red and blue regions regions respectively) appear to  have maximum values centered on the orbital/spin phases $\sim$0.4/0.65 and $\sim$0.4/0.15 respectively. These coincide with the maximum peaks in the photometry and the centers on the ``split" linear polarization peaks. Comparing the left and right third panels also reveals that the orbit, beat image has the diagonal structures of similar orientation to those present in the photometry and linear polarization. However, in comparison, the orbit, spin circular polarization structures appear to be more vertically orientated suggesting that the circular polarization is spin dominated.

The bottom two panels of Fig. 2 show the simultaneous position angle of linear polarization, also phase-folded and binned in the same manner as the photometry, linear and circular polarization of Fig. 2. The colour coding represents position angle. The orbit-spin image shows vertical features consistent with spin dominated position angle variations, i.e. lines of constant position angle run along lines of constant spin phase. The position angle appears to rotate twice through 180 degrees over the course of one spin cycle. The orbit-beat image (right panel of Fig. 2) confirms the spin dominance of position angle by exhibiting diagonal lines of constant angle: these would otherwise appear vertical if the position angle varied at the beat period.

\begin{figure}
    \centering
	\includegraphics[width=\columnwidth]{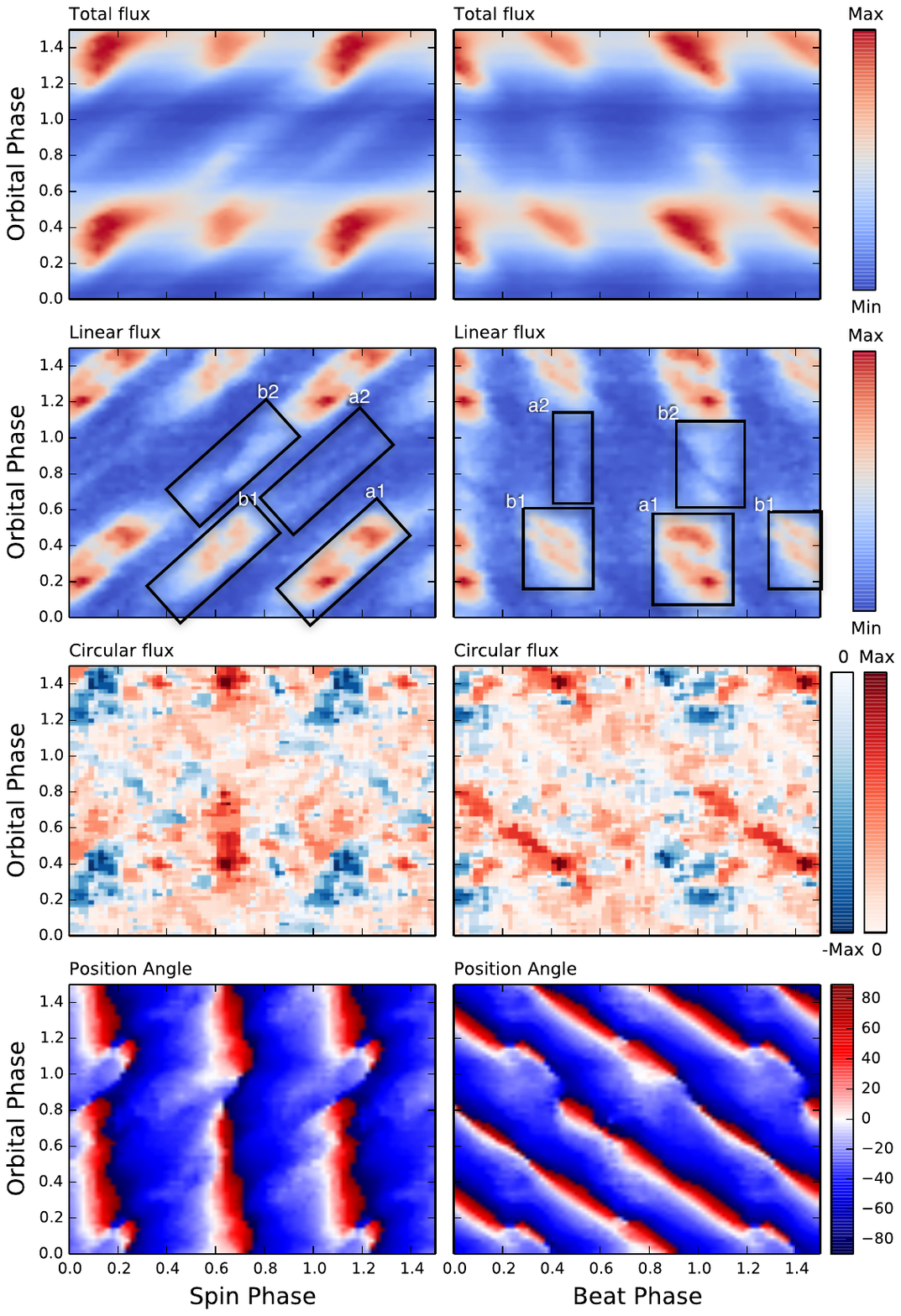}
    \caption{Dynamic pulse profile images showing how the spin and beat modulations of total intensity and polarization parameters change with orbital phase. These represent the averaged phase-folded data from 14, 25, 26, 27 May 2016 and 27, 28 March 2017, showing (from top to bottom), total flux, linearly polarized flux, circularly polarized flux and position angle of the linear polarization vector. The linearly polarized flux has been normalised, while circularly polarized flux has not. See text for description of labeled rectangles.}
    \label{fig:example_figure}
\end{figure}

\section{Fourier Analysis}

We subjected the entire time series of linearly polarized flux to a Fourier analysis and present the amplitude spectrum of this data in Fig. 3. All of the peaks in the amplitude spectrum were able to be identified with the known periodicities. Multiple harmonics of the spin ($\omega$) and beat ($\omega-\Omega$) frequencies are clearly identified as well as the spin$-$orbit sideband combinations of the spin and orbital frequencies, which are expected to occur at frequencies $\nu = m.\omega \pm n.\Omega$, where {\it m \& n} are integers. We see that the spectrum is heavily aliased due to daily, weekly, monthly and yearly gaps in the dataset and is very similar to the photometric amplitude spectra presented in \cite{Potter2018} for the total intensity, {\it I}.

\cite{Potter2018} described a technique of searching all the amplitudes of the spin, orbit and sidebands variations and harmonics as a function of spin and orbit frequency parameter space in order to decide which are the most likely periods. Despite there being several aliases, not all of them have corresponding amplitude peaks at the various spin/orbit/sideband frequencies and harmonics, resulting in a reduced ``summed amplitude''. See \cite{Potter2018} for further details and how it was applied to the photometric observations. This technique allows all possible information on the periodicities, based on the identification of the two fundamental $\omega$ and $\Omega$ frequencies, their sidebands and harmonics, to be combined to produce the highest signal-to-noise and best resolution period peak in the combined amplitude spectrum. The frequency resolution is then defined by the highest frequency signal present in the data set, which is the 4th harmonic of the spin or beat periods.

In Fig. 4 we show the results of the technique applied to our linear polarization data. The tallest peak is consistent with the tallest photometric peak from \cite{Potter2018}, occurring at the beat frequency. There are other smaller but comparable peaks located at the $\sim$1 year aliases, however \cite{Potter2018} have shown that these are significantly reduced in the similar photometric analysis that consisted of a longer dataset. Furthermore we find that the polarimetry does not require a spin-derivative to correctly phase our datasets over the 2 year time base as was found for the photometry \citep{Potter2018}.

\begin{figure}
	\includegraphics[width=\columnwidth]{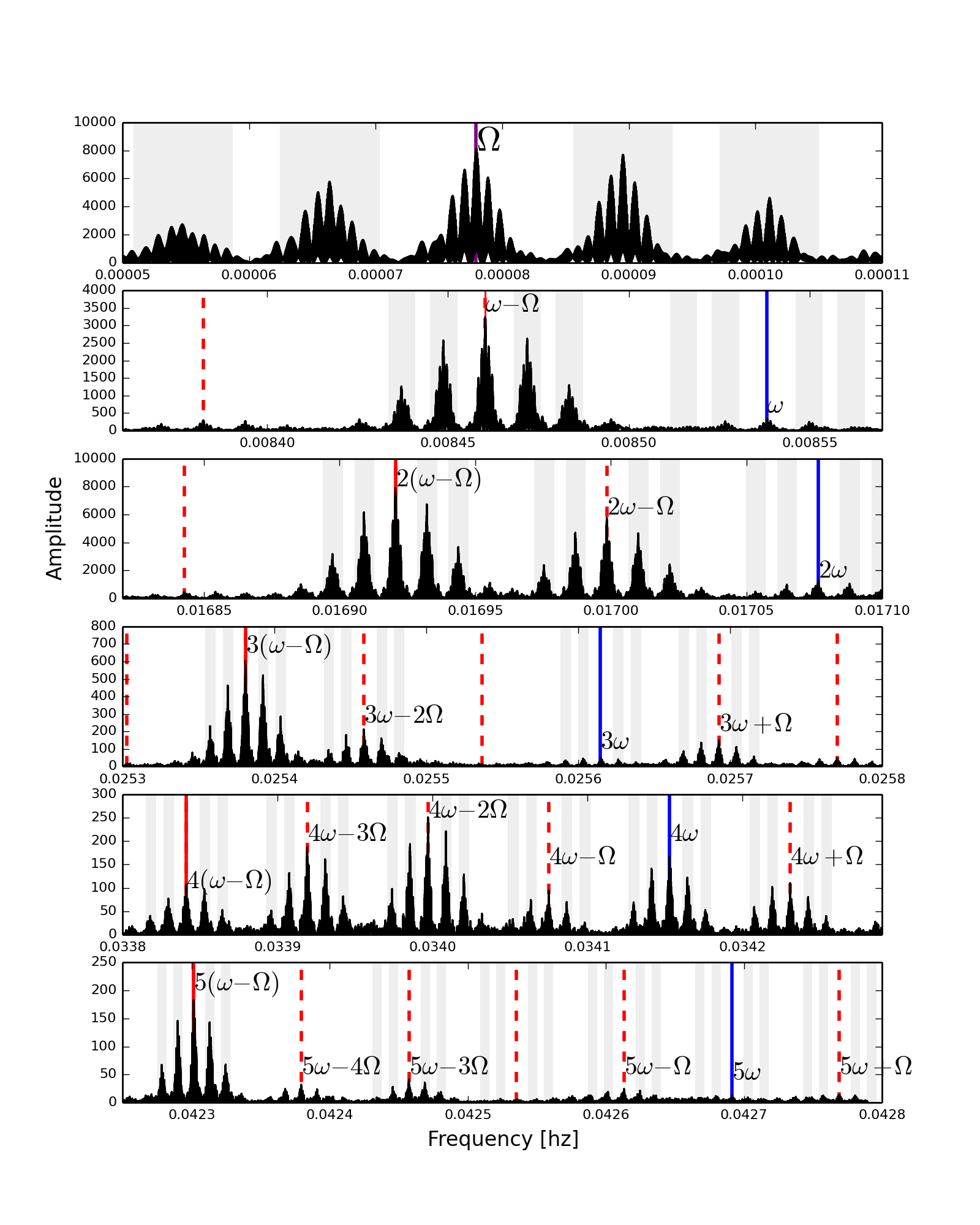}
    \caption{Amplitude spectrum of the total linearly polarized flux obtained with HIPPO in 2016 and 2017. The panels show the amplitude spectrum centered on regions around the orbital, beat, spin and sideband frequencies, plus their harmonics. The blue lines indicate the spin frequency and its harmonics while the solid red lines indicate the beat frequency and its harmonics. The dashed red lines indicate other sideband frequencies. One day aliases are indicated by the vertical grey bars, while the $\sim$15 day aliases are also visible, particularly in the top panel, as separated black period peaks. The amplitude spectrum also contains $\sim$1 year aliasing, which are unresolved in this figure, but are plainly seen in high resolution plots, like in Fig.4. }
    \label{fig:example_figure}
\end{figure}

\begin{figure}
    \centering
	\includegraphics[width=\columnwidth]{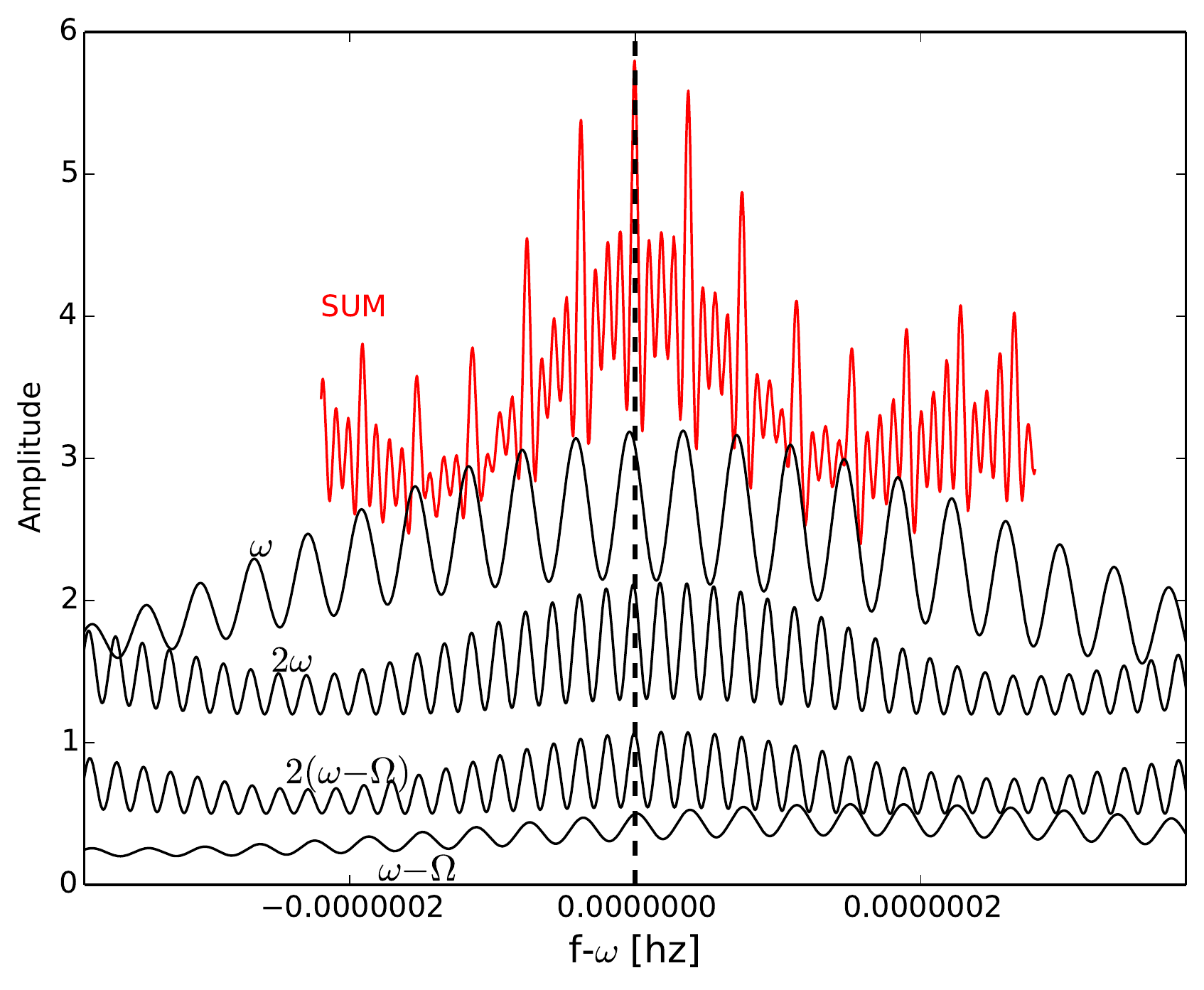}
    \caption{Amplitude spectra for the linearly polarized flux where the black curves are magnified views centered on the spin and beat frequencies and the spin harmonics, re-scaled in frequency space with respect to the spin fundamental (e.g. the frequencies centered on $2\omega$ were divided by 2, etc). Individual peaks represent the 1 cycle/year aliases. These rescaled amplitude peaks (including others identified in Fig. 3 but not plotted here) were then summed to produce the red curve. All spectra have been vertically displaced to aid visualization. The vertical black dashed line indicates the location of the spin and beat harmonics using the center of the tallest peak in the summed red spectrum. It also coincides with the photometric spin frequency determined in \citep{Potter2018}. }
    \label{fig:example_figure}
\end{figure}

\section{The spin modulated polarization}
In order to extract the weaker spin modulated linear polarization signal (e.g. see \cite{Buckley2017}), we pre-whitened our polarimetric time series to remove the effects of the stronger beat modulation. This was done by a non-linear least squares fit of multiple sinusoids fixed at the sideband frequencies and their harmonics, determined in the previous section. Only the amplitudes and phases were allowed to vary in the fitting procedure. 

In Fig. 5 we show the results of the pre-whitening, in the same manner as shown in Fig. 2. Removal of the sideband frequencies, predominantly the $\omega - \Omega$ beat frequency, enhances the spin modulation at all orbital phases, albeit at a lower relative level in the range $\phi_{orb}$ = 0.6$-$1.2. The bottom panels show the orbitally average pre-whitened linear polarization intensity curves, for both the spin and beat periods.  The spin folded curve reveals structures that are stable over the orbital phase implying emission region(s) fixed in the white dwarf rotating frame.

\begin{figure}
    \centering
	\includegraphics[width=\columnwidth]{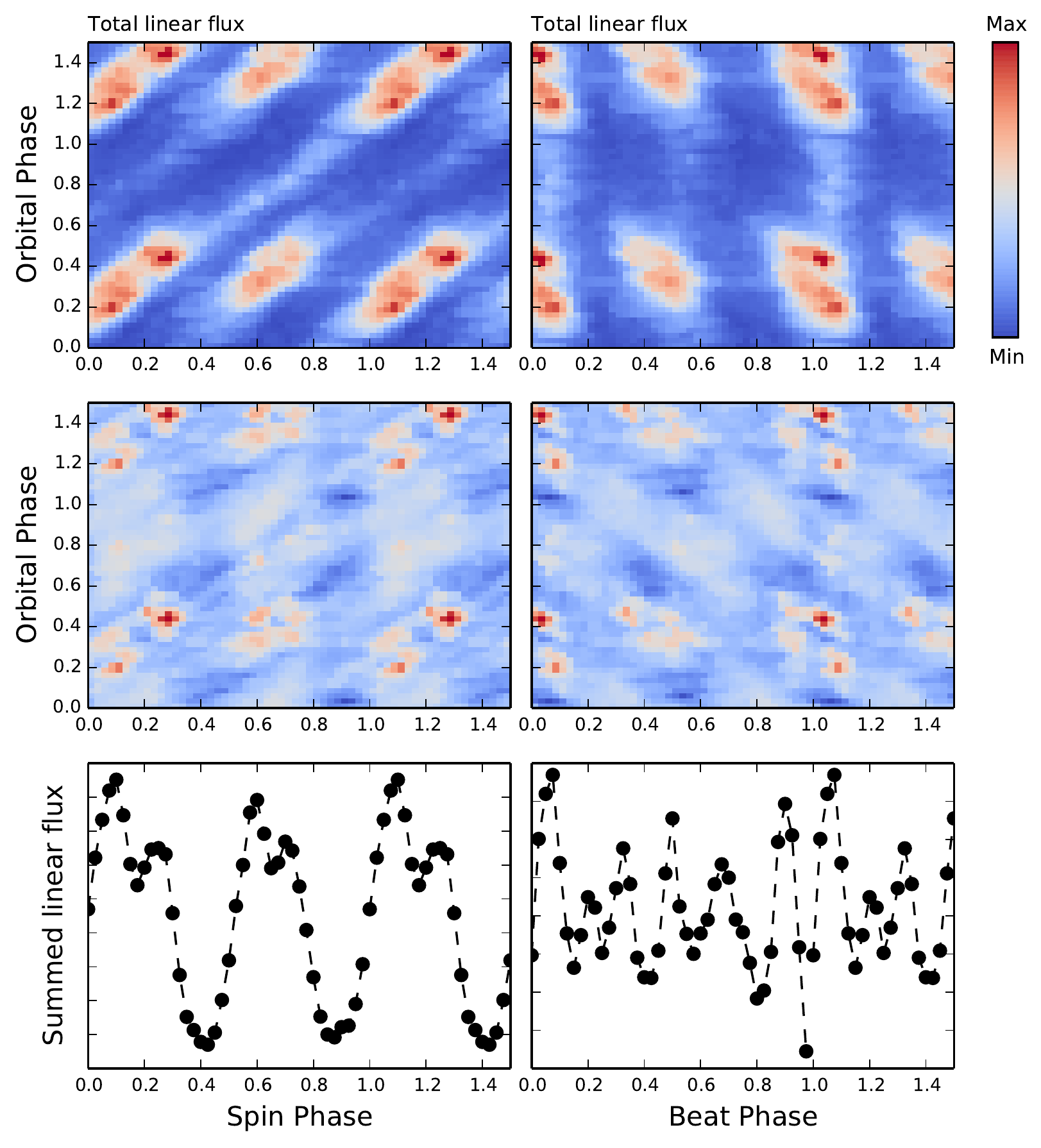}
    \caption{Dynamic pulse profile images showing how the spin and beat modulations of linearly polarized flux change with orbital phase before and after pre-whitening. The top two panels are the un-pre-whitened data, as in the 2nd panels of Fig. 2. For the middle two panels the data was pre-whitened by the beat period and its various harmonics. The bottom two panels show all the data binned on the spin and beat period, summed over all orbital phases.}
    \label{fig:example_figure}
\end{figure}

\section{Proposed model}
\begin{figure}
    \centering
	\includegraphics[width=\columnwidth]{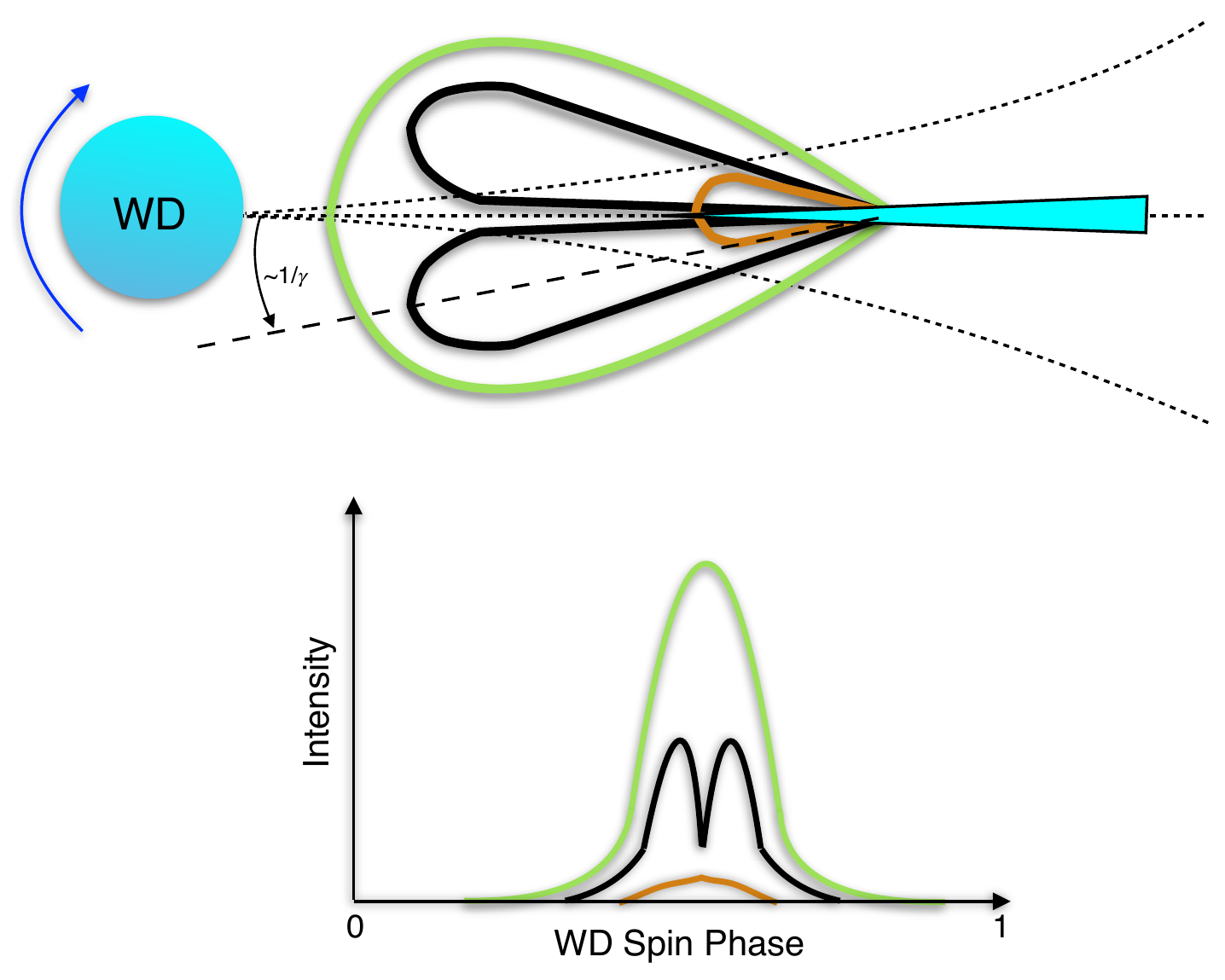}
    \caption{Schematic showing the beamed geometry of synchrotron emission from relativistic electrons (cyan area) approaching the first magnetic mirror point near the magnetic WD. Green, black and brown components represent the expected beamed geometry of the total intensity, linear and circular polarized emissions. Lower plot shows the corresponding white dwarf spin-pulse profiles. Not to scale.}
    \label{fig:example_figure}
\end{figure}

We consider a simple model, based on geometrical components only, in order to explain the spin, beat and orbital modulations of the polarized emission which is understood to be of synchrotron in origin e.g. \cite{Marsh2016}, \cite{Geng2016}, \cite{Buckley2017}, \cite{Katz2017}, \cite{takata2017}. The characteristics of our model for the polarized emission ultimately depends on the location of the synchrotron emission source(s) in the binary frame. We attempted a few models that placed the synchrotron emission source fixed in the binary frame e.g. on the irradiated face of the secondary star or 
a bow shock. We were unable to reproduce the morphology of the polarimetric modulations using such emission sites. Specifically we could not construct a model that reproduces the detailed linear polarized structures labeled as regions a1, a2, b1 and b2 in Fig. 2., the orientation of the circular polarized structures and also, very importantly, the position angle dependence. We would expect to see more vertical structures in the beat dynamic pulse profiles if the emission regions were located at or near the secondary star.


Finally we considered scenarios where the emission regions are locked in the white dwarf rotating frame. One such scenario involves relativistic electrons accelerating from the vicinity of the secondary star \textit{towards} the white dwarf.  \cite{Buckley2017} and \cite{takata2017} have suggested that magnetic interaction on the surface of the companion star heats its stellar surface. Furthermore, \cite{takata2017} proposed that this interaction produces an outflow of electrons with relativistic energies. The electrons are injected and trapped in closed magnetic field lines of the white dwarf. Most of the relativistic electrons are captured by field lines from the magnetic pole that happens to point towards the secondary as it sweeps past it, giving rise to bright synchrotron emission. However, some electrons flow in the opposite direction, along the same magnetic field lines, towards the opposite magnetic pole which also produces polarized synchrotron emission, but at at a lower level. \cite{takata2017} reasoned that the relativistic electrons do not reach the surface of the white dwarf but instead emit synchrotron emission as they reach magnetic mirror points on approach to the white dwarf. These electrons are reflected and may end up being re-captured by the secondary star or accelerating back towards the opposite magnetic pole.

The synchrotron emission from accelerating electrons trapped on magnetic filed lines is depicted in the upper schematic of Fig. 6. The synchrotron emission is ``beamed" in the direction of motion of the relativistic electrons, resulting in total intensity as well as linear and circular polarized pulses as the white dwarf rotates (Fig. 6, lower schematic). The beaming pattern for the linearly polarized emission from relativistic electrons takes on a double peaked profile, whereas the the circularly polarized emission is beamed along the direction of the relativistic electrons. The amount by which the emission is beamed depends on the Lorentz factor ($\gamma$) of the relativistic electrons. 
Observationally, other factors, 
such as the system inclination and the angle of the dipole magnetic field relative to the spin axis of the white dwarf, will determine the observed characteristics of the pulse profile.  As we will now explain, the critical feature of the model is that the synchrotron emission region(s) are locked and rotate with the spinning white dwarf.

For the simple model described above, after some experimentation we chose an inclination of $60^{\rm o}$ and a dipole offset angle of $40^{\rm o}$. We settled on a synchrotron emission profile with a ``beaming angle'' of 45$^{\rm o}$, i.e. the peak in linear polarization is seen at 45$^{\rm o}$ from the magnetic field direction, indicated as $\sim1/\gamma$ (albeit not to scale) in Fig.6. These parameters were iteratively chosen by hand in order to approximately match the pulse profile widths and separations as seen in the observations. A more detailed exploration of the parameter space through model fitting is beyond the scope of this work and we stress that the values chosen here are by no means a unique solution, i.e. the model pulse widths and their separations can be adjusted by tweaking any of the three parameters described above, but the underlining geometrical components and resulting emissions remain the same.

Fig. 7 shows how a data set, produced by our simple model for a single rotating synchrotron source (i.e. not in a binary), would appear when phase-fold-binned in the same manner as in Fig. 2. The left panels show that the spin-pulses trace a vertical unchanging profile as a function of time (where Orbital Phase is a proxy for time in this single-star case). The right hand panels show how these pulses appear when folded on the beat frequency. As expected they take on a diagonal appearance because the spin-modulated pulses appear slightly earlier in beat-phase as a function of time. 

We next took into consideration the right-hand panels of Fig. 2. (first and second panels), which indicate that the brightest pulses occur in the beat-phase range $\sim$0.9$-$0.1 (labeled as region a1). We adjusted our model such that the total synchrotron intensity increases during those beat-phases. This is depicted in Fig. 8 where the synchrotron emission is intensified as the corresponding magnetic pole (pole a) passes through the thick green area.
However, from the observer's perspective, these pulses will appear brighter during orbital phases $\sim$0.2$-$0.45 only, because of the beaming direction of the synchrotron emission. Note that region (a1) in the beat dynamic pulse-profile transposes to region (a1) in the spin dynamic pulse-profile. 

A possible astrophysical scenario for the increased brightness is discussed later. Suffice to say that the beat-phase dependency is a function of the angle between the WD magnetic poles and the M-dwarf. Therefore it is as a result of some astrophysical mechanism within the binary frame and not observer viewing angle. We suggest that the magnetic poles receive an enhanced injection of relativistic electrons as they sweep past the M-dwarf, as suggested by \cite{takata2017}. This subsequently results in an increase in synchrotron emission as the electrons reach a magnetic mirror point close to a magnetic pole of the WD. The beat-phase range in which the increased emission occurs is indicted by the green arc-like region in Fig. 8.

Fig. 9. shows the phase-folded-binned model data, with the beat-phase constraint. Our model is now closer to recreating some of the main features of the observations seen in Fig. 2. 
Specifically within region (a1): the diagonal appearances of the brighter intensity pulses in the spin and beat dynamic pulse profiles (top 2 panels in Figs. 2 and 9), the vertical separation of the double linear pulses in the beat dynamic pulse profile (2nd right panel of Figs. 2 and 9), the diagonal separation of the double linear pulses in the spin dynamic pulse profile (2nd left panel of Figs. 2 and 9), the diagonal and vertical regions of the negative (blue) circular polarization in the beat and spin dynamic pulse profiles respectively (3rd panels of Figs. 2 and 9). In our model the peak in negative circular polarization occurs when pole (a) is most pointing away from the observer during spin-phase $\sim$0.1-0.2, i.e. when the synchrotron beaming is most pointing towards the observer.

So far we have considered the emission from a single synchrotron emission region associated with one magnetic pole only. As already indicated by \cite{Marsh2016} and \cite{Buckley2017}, Pulsar-like particle acceleration, the observations are consistent with optical emission from two synchrotron emission regions. Therefore we have added a second, fainter, synchrotron emission region, associated with the second magnetic pole (pole b) which is diametrically opposite the first pole, as depicted in Fig. 8. Similar to the first pole we added the constraint of increasing the intensity of its emission when the magnetic pole sweeps through the region in the binary frame indicated by the same thick green area in Fig. 8. This occurs over the beat-phase interval $\sim$0.3$-$0.55 (as defined by the position of pole (a) in the binary frame) and results in the emission region (b1) in Fig. 2. 

Fig. 10 shows the beamed polarized emission from our two synchrotron emission regions displayed as dynamical pulse profiles in the same manner as the previous figures. The addition of the second region has effectively duplicated a fainter version of the double pulsed emissions from the first region but with an offset consistent with being diametrically opposite. The peak of positive circular polarization coincides with when the magnetic pole (b) is most pointing away from the observer during spin-phase $\sim$0.65, i.e. when the beamed emission is most pointing towards the observer. However the positive circular emission is most enhanced when pole (b) passes through the green region at beat phase $\sim$0.3$-$0.55 and most pointing towards the observer at orbital phases$\sim$0.2$-$0.45 (see 3rd  panels of Figs 2 and 9).

Note that we have artificially adjusted the relative total intensities of the two synchrotron emission regions in order to match the observations. Intrinsically the beamed intensities of both synchrotron emission regions could be roughly equal. However a combination of system inclination and dipole offset angle will result in one pole preferentially beamed in the direction of the observer.

The bottom two panels of Fig. 10 show the model position angle of linear polarization, also phase-fold-binned in the same manner as the total flux, linear and circular polarized flux.  The spin-dynamic pulse profile shows vertical features consistent with spin dominated position angle variations as seen in the observations (Fig. 2), i.e. lines of constant position angle have a constant spin phase. The position angle appears to rotate twice through 180 degrees over the course of one spin cycle, irrespective of orbital phase, as one expects from the two emission regions. The beat dynamic pulse-profile confirms the spin dominance of position angle by exhibiting diagonal lies of constant position angle.

From Fig. 2. we have identified additional pulses not yet accounted for in the model and are labeled as regions (a2) and (b2). These pulses are seen during orbital phases $\sim0.6$-$1.1$. We find that the simplest explanation for the pulses is that they are associated with additional beamed emission from the smaller fraction of electrons that have flowed in the opposite direction 
along the same magnetic field lines, towards the opposite magnetic pole. For example, in Fig. 8. the beamed emission (associated with region a1) is depicted by the green lobe and is best viewed during orbital phases $\sim 0.2-0.45$, as already discussed above. We have also added beamed emission from the opposite magnetic pole, indicated by the fainter green lobe near pole (b). As can be seen from Fig. 8, the associated beamed emission is best viewed at orbital phases $\sim 0.7$-$0.95$ and is associated with region (b2) in Fig. 2. Effectively each magnetic pole, labeled as (a) and (b), receives two injections of relativistic electrons. Injection (1) occurs when the magnetic pole sweeps past the secondary and injection (2) when it is diametrically opposite.

Therefore, similar to above, region (b1) in Fig. 2. is as a result of pole (b) passing through the green enhancement region where the beamed emission is from the first magnetic mirror point. Some relativistic electrons will also travel to the opposite magnetic pole, this time associated with pole (a) where the emission appears in region (a2) in Fig. 2.

Furthermore, a close inspection of the beat dynamic pulse-profile (Fig. 2. 2nd, right panel) reveals a diagonal narrow ``dark'' band running from the bottom right to the top left, linking regions (b1) and (b2). This transposes to the vertical band, centered on spin phase $\sim$0.65, in the corresponding spin dynamic pulse-profile and is understood as the linear intensity dip between the linear pulses (see Fig. 6). We have not emulated the second electron injection in our model, but it is clear from Fig. 10 that the underlining pulse emissions (and dark, narrow bands) are consistent with this scenario. 

Outside of the electron enhancement region (thick green region in Fig. 8), a minimal level of emission is expected to continue as electrons are trapped between the magnetic mirror points. In particular, looking at Fig 8., our model predicts that during beat phase ranges $\sim$0.1$-$0.3 and $\sim$0.6$-$0.85,  neither magnetic pole is passing through the enhancement region. This is verified in the observations which show dark vertical bands (especially the linear flux) during those phases in the beat-pulse dynamic profile (Fig. 2, second right panel). These are transposed to the diagonal dark bands in the corresponding spin-pulse dynamic profile (Fig. 2, second left panel).

We have have shown that the scenario of electrons injected from the secondary star into the magnetosphere of the white dwarf would be consistent with our geometrical model. However this does not preclude other scenarios in which the synchrotron emission region(s) are fixed and rotate with the magnetosphere of the white dwarf. For example, as noted by \cite{Buckley2017}, pulsar-like particle acceleration can occur as a result of $\sim 10^{12}V$ electric potentials between the white dwarf and the light cylinder. They also argue that striped relativistic magnetohydrodynamic wind, outside the light cylinder, may also be present which will also result in pulsar-like particle acceleration. As long as the emission sites are fixed in the white dwarf rotating frame then these scenarios could also be consistent with the observations. Our model depicted in Fig. 8. used an enhanced injection of electrons from the secondary to re-produce the detailed spin, beat and orbital modulations. The orbital phase range at which the maximum amplitude of the polarized pulses are observed is also consistent with this scenario. The other scenarios would resemble something similar to Fig. 11 and would require some other astrophysical mechanism to produce the detailed polarimetric modulations. The fact that AR Sco is a binary system suggests that the secondary star somehow influences the electric potentials responsible for particle acceleration.




\begin{figure}
    \centering
	\includegraphics[width=\columnwidth]{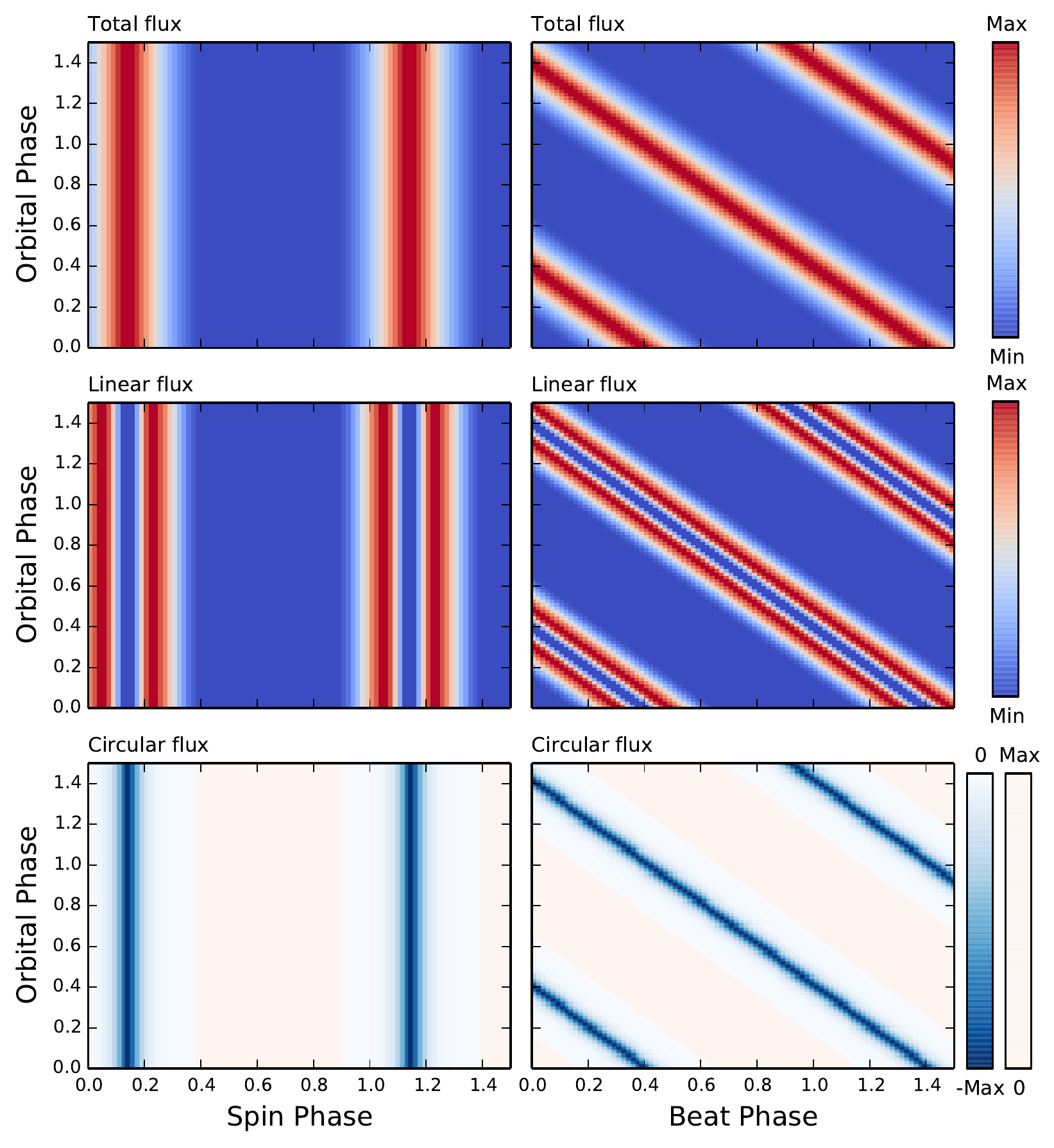}
    \caption{Model simulations of the dynamic pulse profiles of the spin and beat modulated polarization parameters as a function of orbital phase for emission from one magnetic pole of a single spinning magnetic white dwarf (in this case orbital phase is meaningless and is a proxy for time). }
    \label{fig:example_figure}
\end{figure}

\begin{figure}
    \centering
\includegraphics[width=\columnwidth]{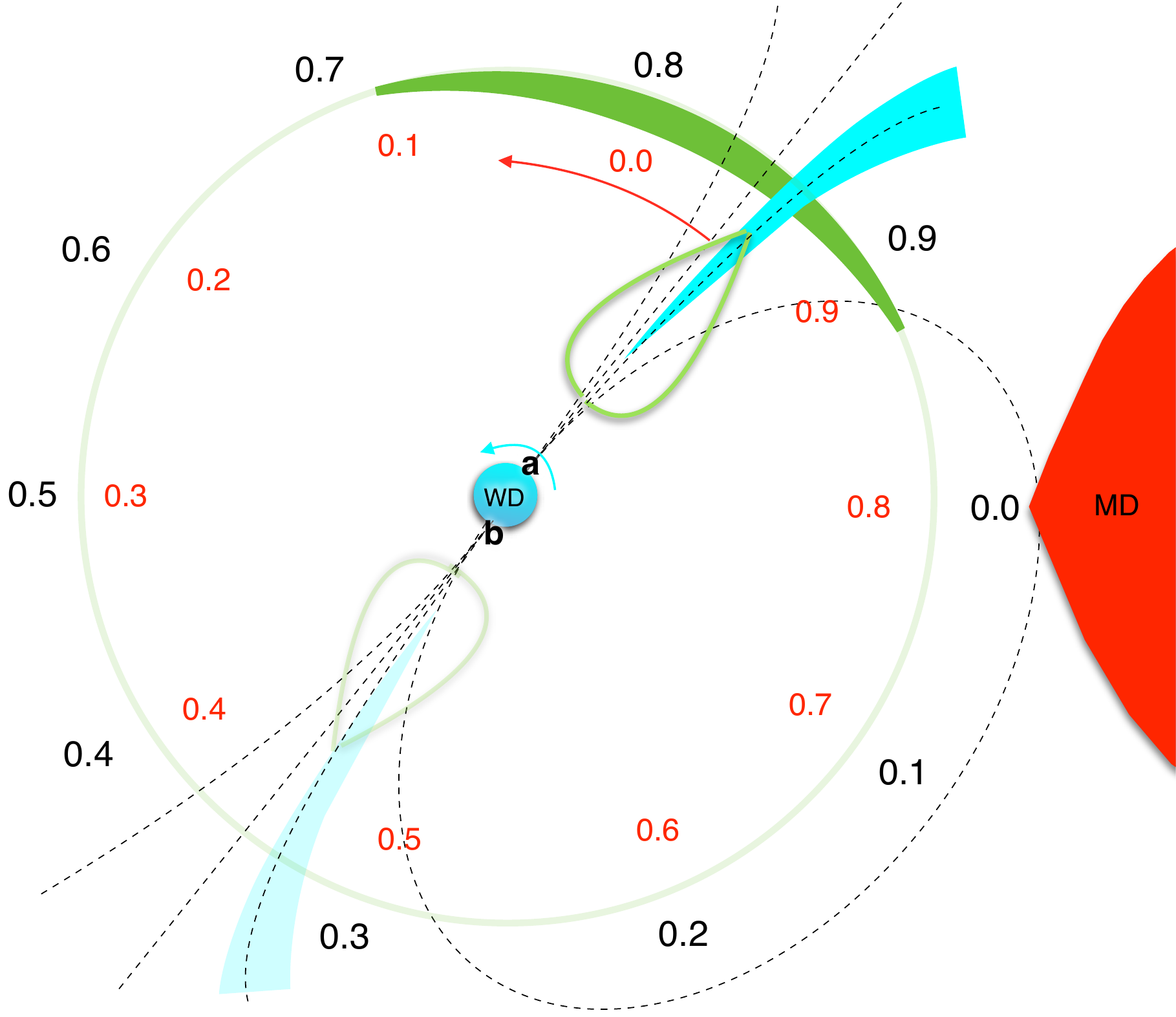}
    \caption{Schematic of the AR Sco system (not to scale). Orbital phases marked as black digits and indicate direction to observer. Beat phases marked as red digits and indicate the angle of the magnetic pole associated with the brighter synchrotron emission region with respect to the M-dwarf. Dashed lines indicate magnetic field lines from close to the two magnetic poles. Thickening of the green circle indicates the beat-phase range in which synchrotron emission is intensified. Cyan region indicates the region where relativistic electrons are nearing a magnetic mirror point giving rise to beamed synchrotron emission (green lobes). Pulses from the fainter synchrotron emission region will also increase in intensity as the associated magnetic pole sweeps through the green region. The green circle is fixed and rotates with the binary frame. }
    \label{fig:example_figure}
\end{figure}

\begin{figure}
    \centering
\includegraphics[width=\columnwidth]{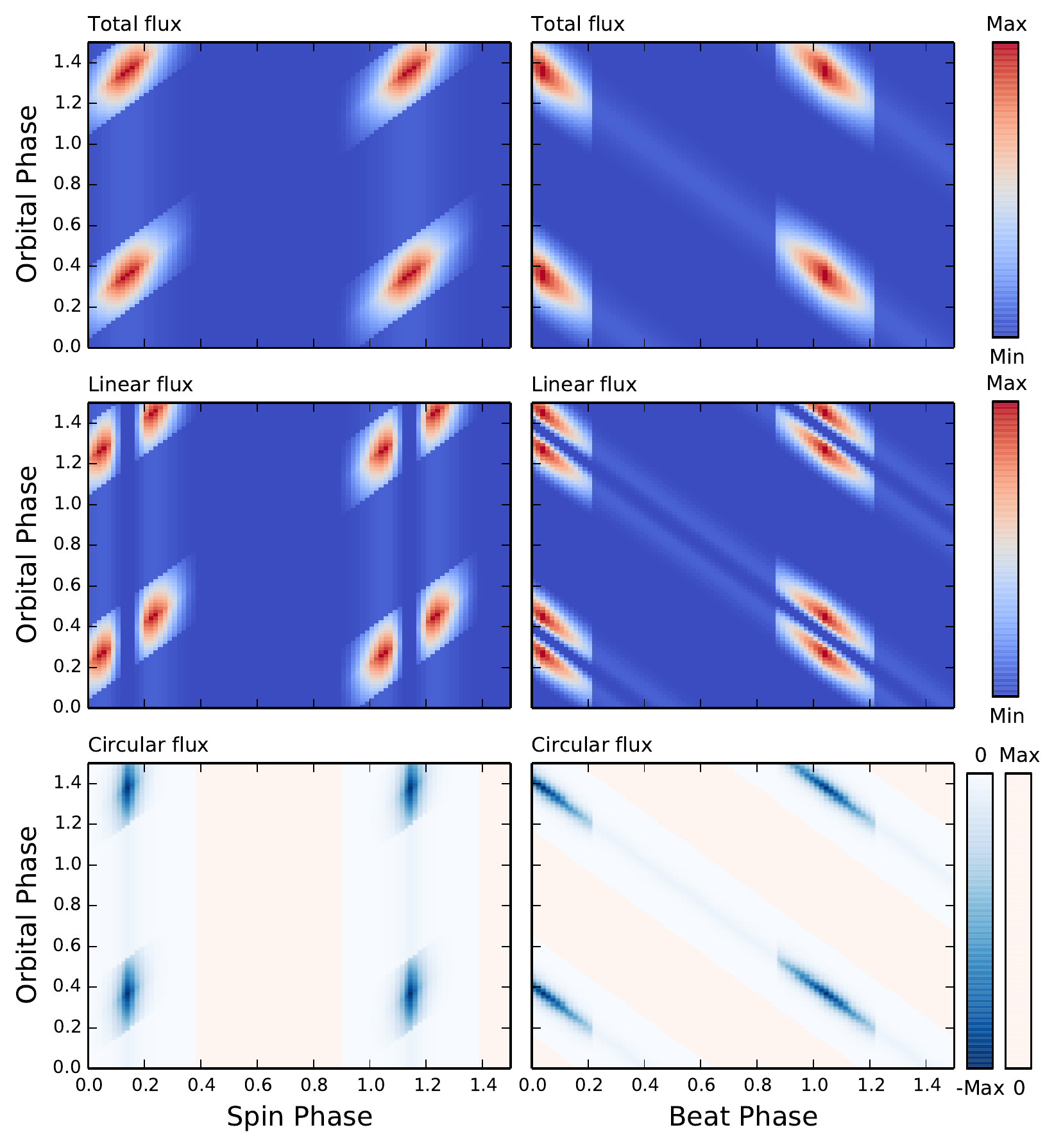}
    \caption{Model simulations of the dynamic pulse profiles of the spin and beat modulated polarization parameters as a function of orbital phase for emission from one magnetic pole of a spinning magnetic white dwarf in a binary system whose companion modulates the polarized emission.}
    \label{fig:example_figure}
\end{figure}

\begin{figure}
    \centering
	\includegraphics[width=\columnwidth]{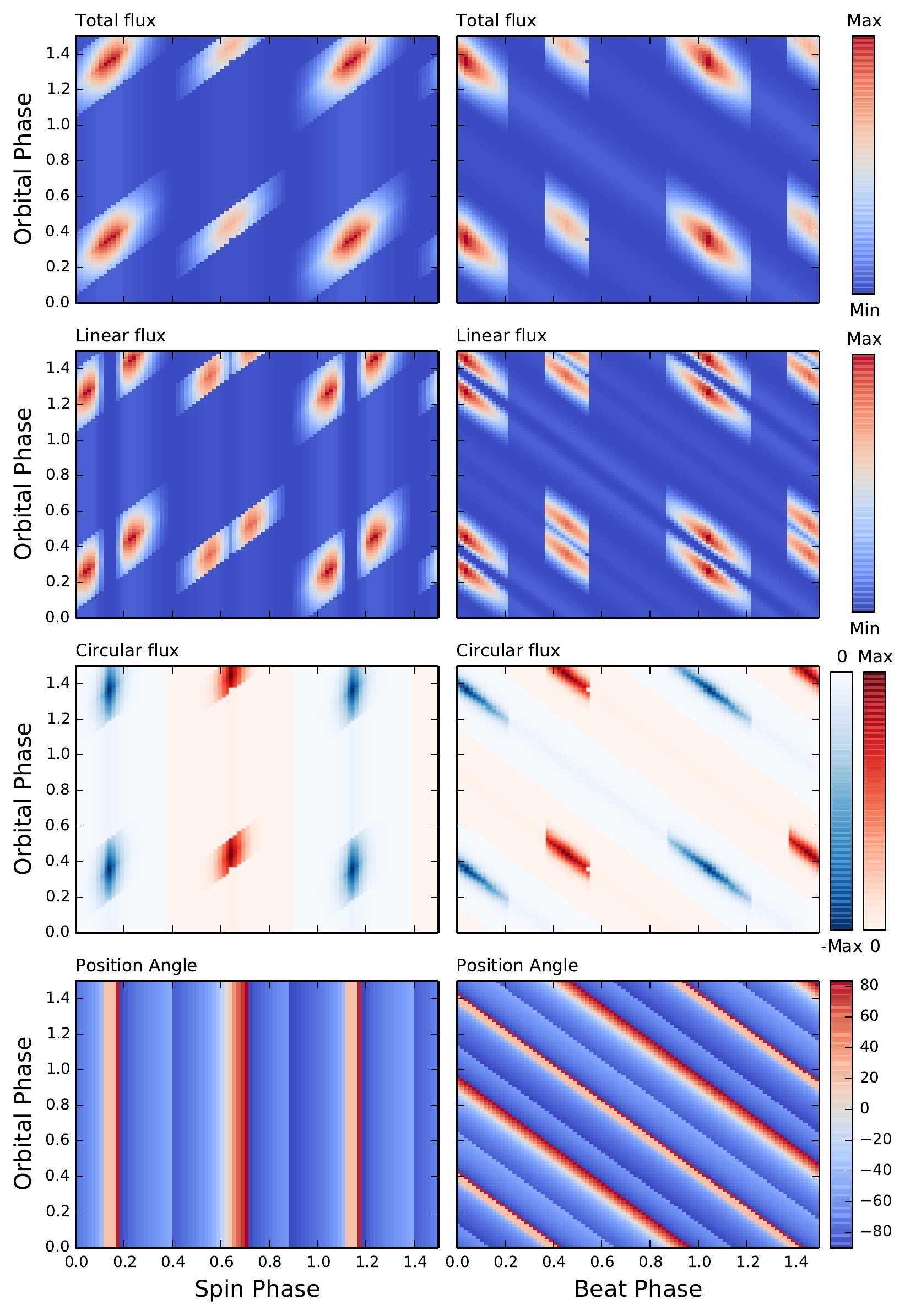}
    \caption{Model simulations of the dynamic pulse profiles of the spin and beat modulated polarization parameters as a function of orbital phase for emission from two magnetic poles of a spinning magnetic white dwarf in a binary system whose companion modulates the polarized emission.}
    \label{fig:example_figure}
\end{figure}

\begin{figure}
    \centering
	\includegraphics[width=\columnwidth]{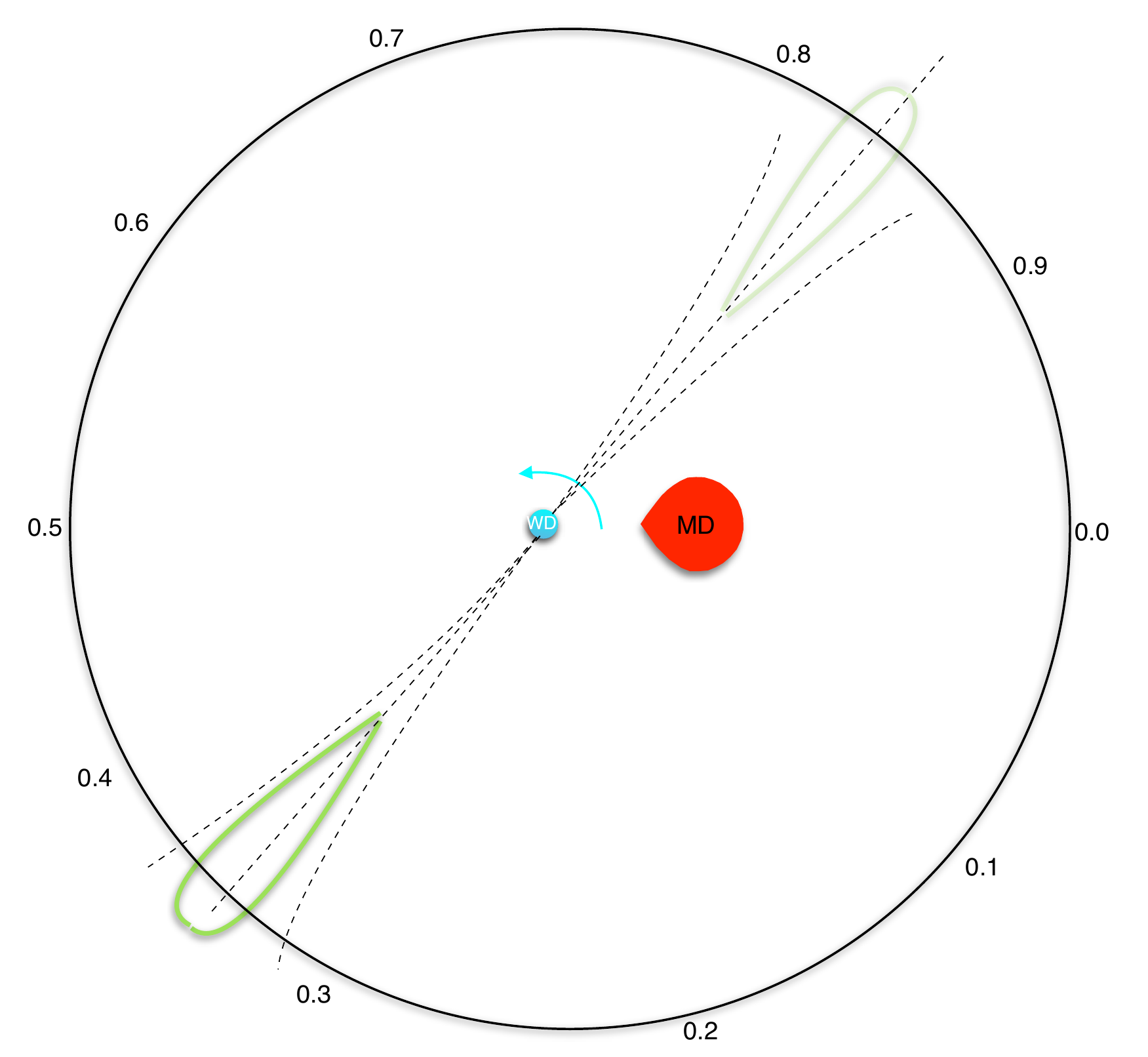}
    \caption{Schematic of the AR Sco system (not to scale). Orbital phases marked as black digits and indicate direction to observer. Dashed lines indicate magnetic field lines from close to the two magnetic poles. The beamed synchrotron emission sites are locked in the white dwarf rotating frame near radius of the light cylinder.}
    \label{fig:example_figure}
\end{figure}

\section{Summary and discussion}
\subsection{Spin, Beat and Orbital polarimetric modulations}
We present $\sim$65 hours of high-speed photo-polarimetric observations, spread over two successive observing seasons, of the recently discovered white dwarf pulsar AR Scorpii. Our observations confirm the highly-polarized emission originally reported by \cite{Buckley2017}, from data with only partial orbital coverage. It also shows definite detection of circular polarization, of both polarities, at a level of $\sim \pm 3$ percent, albeit only briefly at spin phase 0.1 and 0.6. Our extensive dataset covers multiple full orbits on several occasions over almost two years. The observations reveal that the polarized emission from AR Sco is remarkably repeatable between our datasets. Specifically, the amplitudes of the polarized spin and beat pulses are clearly modulated as a function of orbital phase. The corresponding Fourier analysis reveals amplitudes at the orbital frequency and multiple combinations of the spin and beat pulses and their harmonics. 

\subsection{Geometrical synchrotron model}
We present a model to explain the modulations of the polarized emission. The model assumes that all the polarized emission emanates from two diametrically opposed magnetic poles from the spinning white dwarf (WD). The amplitude of the two spin pulses are modulated as a function of the relative angle between the magnetic poles and the secondary star. Consequently the observer will measure larger spin pulses at certain orbital phases leading to spin, beat and orbital modulations. This model is not only consistent with the observed polarized spin and beat intensities, but also reproduces the position angle variations as a function of spin and beat phase.

As pointed out by \cite{Geng2016}, the number of particles required to emit the observed pulsed optical emission is significantly larger than can be supplied by the WD itself. However, \cite{takata2017} discuss a possible MHD scenario for modulating the synchrotron emission. They argue that magnetic dissipation/reconnection on the M-type star surface, as previously advocated in \cite{Buckley2017}, heats up the plasma to a temperature of several keV. This leads to the acceleration of the electrons to relativistic speeds, some of which are trapped in the WD's closed magnetic field lines. The observed pulsed component is then explained by the synchrotron emissions from electrons approaching the first magnetic mirror point, which also evolves with orbital phase owing to the effect of the viewing geometry. Our model is consistent with this scenario and, in addition, can account for a second group of fainter pulses due to emission emanating from a smaller fraction of injected electrons reaching the mirror point for the opposite magnetic pole of the WD.

Whilst this scenario reproduces the detailed polarimetric modulations, other scenarios involving emission sites locked in the white dwarf rotating frame are also a possibility. These would require some other astrophysical mechanism to drive the detailed modulations. In may be possible that both the white dwarf and the M-dwarf are sources of charged particles which are accelerated and therefore multiple mechanisms are contributing to the total polarization.

\subsection{Radio emissions}
As discussed in the introduction, \cite{Stanway2018} argued that unlike the optical emission, the radio emission shows weak linear polarization but very strong circular polarization and infer the probable existence of a non-relativistic cyclotron emission component, which dominates at low radio frequencies. Given the required lower magnetic fields, this likely arises from or near the M-dwarf and is therefore not co-located with the optical synchrotron emission.

\subsection{UV and X-ray emissions}
\cite{takata2018} reported that the UV/X-ray emission observed with ({\it XMM-Newton}) also shows orbital, beat and spin modulations and their intensity maximum is located at the superior conjunction of the M-dwarf star's orbit. These properties are naturally explained by the emission from the M-dwarf star and is consistent with a magnetic dissipation/reconnection process on the M-dwarf star's surface that heats up the plasma to several keV, which also accelerates the electrons to relativistic speeds.

\section*{Acknowledgements}
This material is based upon work by the authors which is supported financially by the National Research Foundation (NRF) of South Africa. 









\bsp	
\label{lastpage}
\end{document}